# Thermodynamics of site-specific small molecular ion interactions with DNA duplex: a molecular dynamics study

Soumadwip Ghosh, Mayank Kumar Dixit, Rajarshi Chakrabarti[*]

[*]Email: rajarshi@chem.iitb.ac.in

*Department of Chemistry, Indian Institute of Technology, Powai, Mumbai – 40076, India.*

**ABSTRACT:** The stability and dynamics of a double-stranded DNA (dsDNA) is affected by the preferential occupancy of small monovalent molecular ions. Small metal and molecular ions such as sodium and alkyl ammonium have crucial biological functions in human body, affect the thermodynamic stability of the duplex DNA and exhibit preferential binding. Here, using atomistic molecular dynamics simulations we investigate the preferential binding of metal ion such as $Na^+$ and molecular ions such as tetramethyl ammonium ($TMA^+$) and 2-hydroxy-N,N,N-trimethylethanaminium ($CHO^+$) to double stranded DNA. The thermodynamic driving force for a particular molecular ion-DNA interaction is determined by decomposing the free energy of binding into its entropic and enthalpic contributions. Our simulations show that each of these molecular ions preferentially binds to the minor groove of the DNA and the extent of binding is highest for $CHO^+$. The ion binding processes are found to be entropically favourable. In addition, the contribution of hydrophobic effects towards the entropic stabilization (in case of $TMA^+$) and the effect of hydrogen bonding contributing to enthalpic stabilization (in case of $CHO^+$) have also been investigated.

## I. INTRODUCTION:

Ever since the discovery of the double helix structure of DNA (dsDNA) by Watson and Crick, it has been the centre of attraction for many researchers. A dsDNA molecule stabilized by hydrogen bonds between nucleotides and base-stacking interactions among aromatic nucleobases [1] plays an extremely important role in biological information storage [2]. Due to the polyanionic nature of nucleic acids, they are surrounded by solvents with high dielectric constants (such as water) and positively charged ions (DNA counterions) for conformational relaxation and charge neutralization respectively [3]. Molecular ions are known to possess preferential binding affinity to the grooves and backbones of the DNA and thus affect the overall DNA dynamics and stability enormously. It is well known that both $Na^+$ and $CHO^+$ are biologically relevant in human body. $CHO^+$ serves as a neurotransmitter and methyl-donor in various biological processes [4,5] while sodium as a nutrient is necessary for regulation of blood and body fluids, heart activity and certain metabolic functions [6]. Hence, the present work which offers a detailed understanding of the thermodynamic aspects of small molecular ion binding to DNA duplex through a variety of interactions may be helpful in designing novel DNA functional materials and DNA nano-devices whose efficiency can be controlled by the stability of duplex DNA in the presence of these ions.

It has been found that a combination of electrostatic attraction and the size compatibility of the hydrated monovalent or divalent cations govern the selective binding of these cations to the DNA duplex [7,8]. On the other hand, small molecular ions such as tetramethyl ammonium ($TMA^+$) and 2-hydroxy-N,N,N-trimethylethanaminium ($CHO^+$) bind to the specific regions of DNA duplex, triplex and quadruplex by forming additional hydrogen bonds between molecular ion sites and electronegative DNA base atoms and other non-covalent interactions [9,10] contribute as well. DNA duplexes may also undergo rapid structural transitions in response to certain external stimuli, such as pH [11] of the solution, electrical signals [12] and macromolecular assemblies [13]. The stability of DNA double helix also depends on the relative content of the constituent adenine (A), thymine (T), guanine (G) and cytosine (C) bases. In physiologically relevant buffered solutions, G-C base pairs are more stable than A-T rich base pairs [14]. However, this observation gets reversed when dsDNA is exposed to molecular ions such as $TMA^+$ and $CHO^+$. For example, alkylammonium ions are known to selectively bind to the A-T base pairs in the minor groove of the DNA and thus stabilizing the A-T rich DNA regions over the G-C rich ones [15]. Recently Tateishi-Karimata et al. analyzed ultraviolet melting point curves to demonstrate that A-T base pairs are more stable than G-C base pairs in the hydrated ionic liquid of choline dihydrogen phosphate [16]. This is consistent with the earlier experimental findings by Falsenfeld et al [17]. Chandran et al. used a combination of molecular dynamics simulations and techniques such as circular dichroism, UV-visible spectroscopy and fluorescent dye displacement assay to demonstrate that hydrated ionic liquid cations can penetrate the DNA grooves and influence its thermodynamic stability via hydrophobic effects and electrostatic interactions [18]. Effect of varying $TMA^+$ concentrations on the thermostability of different DNA sequence was observed by Riccelli et al [19]. Portella et al. used MD and NMR analyses to probe the binding of $TMA^+$ and $CHO^+$ to the A-T rich minor grooves of the DNA [20] which is distinctly different from that previously reported for alkali and alkaline earth metal ions [21]. Among other works, Sugimoto & co-workers investigated the importance of solvent-accessible surface area in determining DNA triplex stabilization or destabilization upon binding with $Na^+$, $TMA^+$ and $CHO^+$ [22]. The unique mode of stabilization of the A-T rich tracts in DNA duplex upon choline binding due to the narrower width and electrostatically polar environment of the groove has been studied by the same research group [23]. Among other important works on DNA-molecular ion binding, the early stage of intercalation of doxorubicin (an anti-cancer drug) to two 6 base-pair DNA fragments was observed by Lei et al [24]. Fujiwara & Co-workers demonstrated the effects of metal ions on the conformational difference between two modified nucleotides dGMP and 8-oxo-dGMP which are two major sources of spontaneous mutagenesis [25].

In this paper we attempt to investigate the thermodynamics of preferential ion binding to the grooves and backbones of a dsDNA in order to get a more transparent picture of small molecular ion binding to DNA duplex. We primarily address two issues. Firstly, using all-atom molecular dynamic simulation, we show that all the three species in our study ($Na^+$, $TMA^+$ and $CHO^+$) prefer to bind to the DNA minor groove than the backbones or the major groove and among these three species $CHO^+$ possesses the highest affinity for binding. This conclusion is drawn from $\Delta G_{binding}$ which is computed from the potential of mean force (PMF) for each ion. Secondly, $\Delta G_{binding}$ for each species is decomposed to obtain relative contributions of different thermodynamic driving forces to ion binding to DNA minor grooves. Our simulations show that the negative $\Delta G_{binding}$ for $Na^+$ and $TMA^+$ is supplemented mostly by large gain in entropy, the enthalpic contribution being highly unfavourable due to the breaking of native hydrogen bonds between two parallel DNA helix triggered by molecular ions occupying specific regions of the duplex. In case of $CHO^+$, the ion binding is facilitated by enthalpy as well as entropic contributions since $CHO^+$ is equipped with a polarizable hydroxyl group, capable of forming additional hydrogen bonds with electronegative DNA base atoms [26, 27]. Moreover, it is observed that hydrophobic effects contribute partly to the large entropy gain of the preferential ion binding to DNA minor groves.

## II. DNA MODEL AND SIMULATION DETAILS

All the molecular dynamics simulations are performed using GROMACS 4.5.6 [28]. The all-atom CHARMM force field and potential parameters for nucleic acids [29] are used for the initially generated structure of the DNA duplex with base sequence (5'-CGCGAATTCGCG–3')$_2$, while the explicit SPCE water model [30] is used to solvate the DNA. The PDB file of the canonical B form of Dickerson Drew dodecamer is taken (PDB ID 436D) [31] from Brookhaven protein data bank and the freely available package 3DNA [32] is used for inserting hydrogen atoms. Subsequently, the DNA duplex is kept inside a cubic box containing ~9000 SPCE [33] water molecules (see **TABLE 1**). The structures and topologies of the molecules $TMA^+$ and $CHO^+$ were generated using the SwissParam [34] web service which uses the standard atom types and parameters included in the all atom CHARMM force field [29] directory. The net charge of the native system was found to be -22 and the ions ($Na^+$, $TMA^+$ and $CHO^+$) were added for charge neutralization. Three of these primary systems each containing either 22 $Na^+$ or $TMA^+$ or $CHO^+$ were simulated at three different temperatures (300K, 270K and 330K) in order to investigate the thermodynamics of the DNA-molecular ion binding processes. The protocols and algorithms used for performing molecular dynamics simulations are as follows.

First we employ 1000 steps of steepest descent method [35] in order to lower the potential energy of the system and eliminate any initial stress. Next, the system is equilibrated at constant pressure and temperature (NPT) performed at a temperature of 300K for 5 ns. Once the system reaches the desired pressure, temperature and cell volume, it is equilibrated using the isochoric-isothermal ensemble (NVT) for 2 ns. The NPT step is carried out using Parrinello-Rahman borostat [36] and the v-rescale thermostat [37] is used to keep the temperature of the system constant at 300 K and the system configuration is updated by GROMACS using the leap frog integrator [38]. After the completion of the equilibration steps, the production MD run is started for 60 ns. The entire production MD is carried out with a time step of 2 fs and the information regarding trajectory, velocity and energy are stored after each 1 ps for analysis. The minimum image convention [39] is used to calculate the short ranged Lennard–Jones interactions. The spherical cut-off distance for both electrostatic as well as van der Waals forces is kept at 1 nm. The SHAKE [40] algorithm is used to impose a holonomic constrain on the equilibrium bond distance of the SPCE water molecules. The long range electrostatic interactions are calculated using the particle mesh Ewald [41] method. Since the final production run is performed under NPT condition the PMF obtained as a function of separation between two nanoscopic objects (a particular DNA segment and a molecular or a metal ion) is a direct measure of the Gibbs free energy of DNA- molecular ion association. An overview of the simulated systems (at 300K) in the presence of different ions has been given in **TABLE 1**.

## III. SIMULATION RESULTS AND DISCUSSIONS

| System | V (nm$^3$) | N$_{water}$ | d (kg.m$^{-3}$) | T (K) | P (bar) | [Ion$^+$] (M) |
|---|---|---|---|---|---|---|
| dsDNA-Na$^+$ | 284.665±0.006 | 9176 | 1009.43±0.02 | 300.402±0.004 | 1.056±0.001 | ~0.1284 |
| dsDNA-TMA$^+$ | 280.685±0.001 | 9084 | 1018.55±0.09 | 300.008±0.004 | 1.051±0.001 | ~0.1206 |
| dsDNA-CHO$^+$ | 280.208±0.003 | 9032 | 1020.94±0.01 | 300.008±0.0041 | 1.051±0.0004 | ~0.1301 |

TABLE 1. Overview of the simulated systems. V, d, T, P and [Ion$^+$] represent the cubic box volume, density, temperature, pressure and the ion concentration of the simulated systems, respectively at 300K. N$_{water}$ denotes the number of SPCE water molecules in each of the system. Standard errors are given in the parentheses.

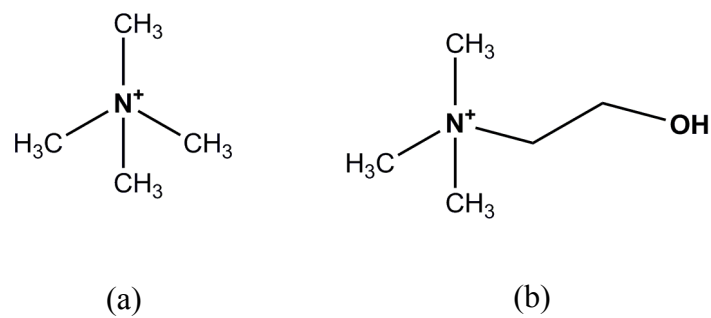

(a)          (b)

FIG 1. Chemical structures of (a) TMA$^+$ and (b) CHO$^+$.

## A. Distribution of molecular/metal ions around the duplex DNA

We calculate the spatial density distribution functions (SDF) for different ions around the time-averaged DNA structure using the g_spatial utility of GROMCAS 4.5.6 [28]. The bright green isosurfaces in **FIG 2** represent the SDFs of the three ions separately around the DNA (represented as van der Waals spheres) helix axis. For the best display, the SDFs of ions are drawn for equal densities > 50 particles/nm$^3$ in each case. It is apparent from **FIG 2** that CHO$^+$ binds most strongly to the central AT-rich DNA minor groove followed by TMA$^+$ and then Na$^+$.

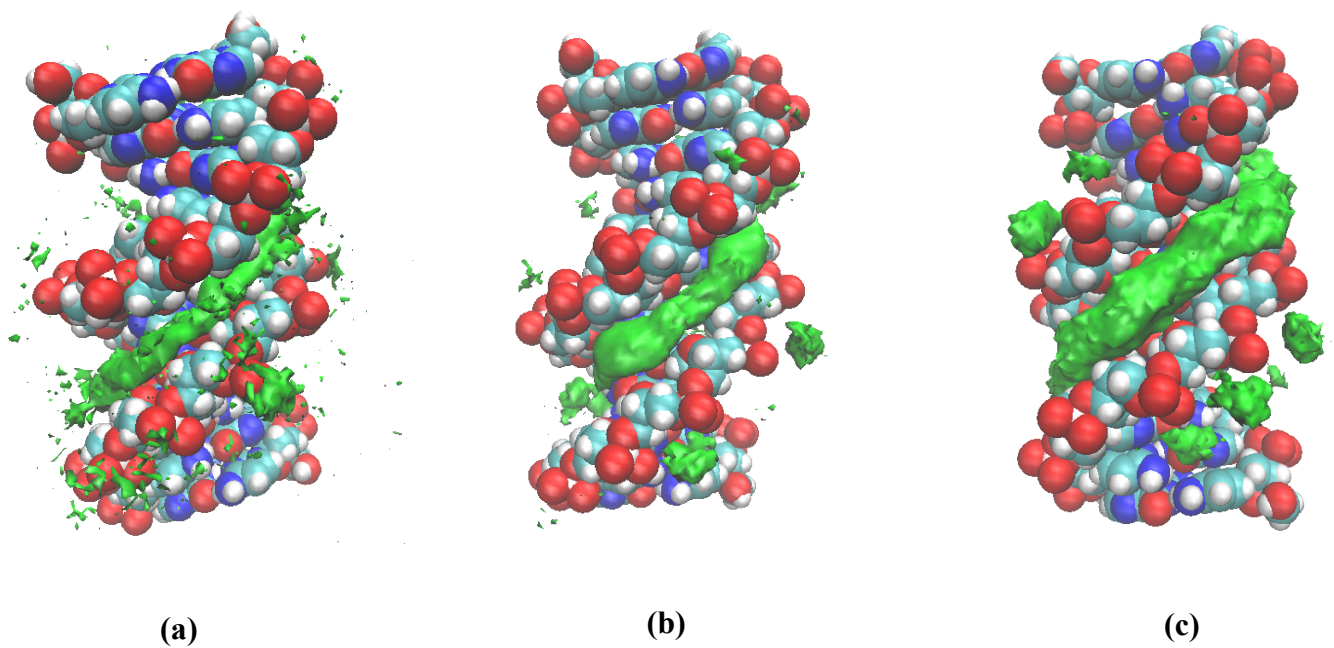

(a)          (b)          (c)

FIG 2. Spatial distribution functions of (a) Na$^+$, (b) TMA$^+$ and (c) CHO$^+$ (bright green isosurfaces) around averaged DNA structure (represented as van der Waals spheres) seen away from the DNA helix axis. Snapshots are rendered using VMD. Water molecules are removed for clarity.

## B. Excess Number of Cations

The calculation of excess number of cations around specific DNA segments [42,43] as compared to that of bulk is of particular interest to us in this study. For each ionic species we calculate the radial concentration profile, $c(r)$ as a function of the distance of the ions from different segments of the DNA according to Robbins et al [44]. In order to compute $c(r)$ for different ionic species we divide the volume around DNA into cylindrical shells $(r, r + \Delta r)$ while r ranges from 0 to $r_{max}$ of 3.2 nm i.e, half of the simulation box length. The number of ions in each shell is averaged over the entire frame of the corresponding MD trajectory. Then we obtain the corresponding cation number density distribution by dividing the number of cations in each shell by the corresponding volume of the cylindrical cell, $2\pi r h \Delta r$, where h is the height of the DNA duplex i.e, ~4.2 nm for the well-known Dickerson Drew dodecamer [44]. The cation number density is then converted to radial molar concentration, $c(r)$ as a function of location of cations within the simulation box.

From a production run, the excess number of a particular type of cation *i* around certain DNA binding site as compared to the bulk is denoted as $N_i$ and it is estimated by integrating the excess ion concentration, $c_i(r) - c_i(max)$:

$$N_i = \int_0^r \{c_i(r') - c_i(r_{max})\} 2\pi r' h dr' \qquad (1)$$

Here, $c_i(r_{max})$ corresponds to the bulk cation concentration in molarity and h is the height of the DNA duplex.

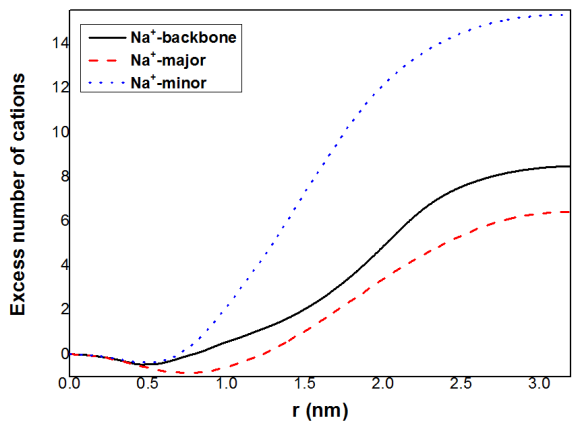

(a)

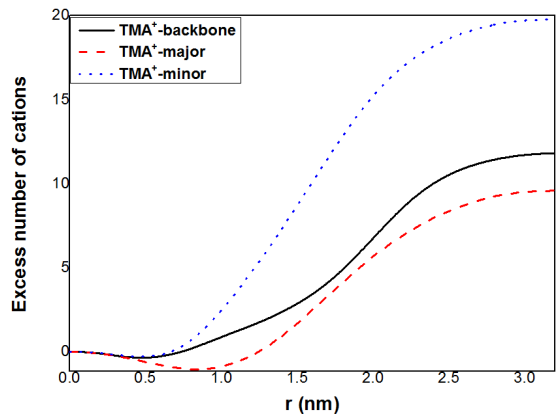

(b)

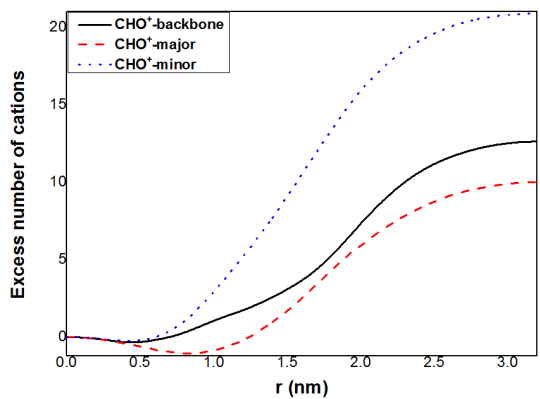

(c)

FIG 3. Excess number of (a) $Na^+$, (b) $TMA^+$ and (c) $CHO^+$ as a function of their separations from DNA backbones, major and minor grooves at 300K respectively.

| $N_p$ | Backbone | Major groove | Minor groove |
|---|---|---|---|
| $Na^+$ | 8.190±0.2133 | 6.101±0.321 | 15.078±0.239 |
| $TMA^+$ | 11.428±0.367 | 9.226±0.352 | 19.453±0.335 |
| $CHO^+$ | 12.350±0.426 | 9.619±0.367 | 20.532±0.368 |

TABLE 2. Excess number of cations ($N_p$) in the plateau regions of Fig 2 corresponding to different binding sites of the DNA at 300K (standard deviations are provided in parentheses).

It is apparent from **FIG 3** that all the three species exhibit higher binding affinity for DNA minor grooves as compared to DNA backbones and major grooves. It is reported recently that the narrower minor groove of DNA duplex consisting of A-T base pairs not only allows multiple hydrogen bonding between $CHO^+$ and DNA but also sustains the native hydrogen bonds between two parallel DNA helices [20]. Our findings are consistent with their observations based on NMR and MD simulations. Our study is also consistent with the DNA minor groove binding of small drug molecules like netropsin and distamycin [45]. It is also worth mentioning that the reasonable convergence of the number of excess cations to that of the bulk (**FIG 3**) correlates well with the work done by Yoo and Aksimentiev [46]. We also tabulate the excess number of cations ($N_p$) in the plateau regions of **FIG 3** that correspond to different binding sites of the DNA for comparison (**Table 2**). This along with the SDFs in **FIG 2** provide primary evidences to the preferential binding of the DNA minor groove atoms with $CHO^+$ over that of $TMA^+$ and $Na^+$. The other measures of the extent of comparative binding among the three species under consideration are determined later based on the potentials of mean force of site specific DNA-small ion binding (**section III.D**)

*C. Two dimensional number density profiles of ions around DNA minor grooves*

Two dimensional number densities of molecular ionic species embedded in the cavity of DNA minor groove can be visualized using the g_densmap utility of GROMACS 4.5.6 [28]. The number density of a particular molecular ion is computed along both the sides of the reference axis formed by connecting the centres of mass of atoms constituting the minor groove in two parallel chains at 300K. Both the axial and radial number densities are plotted in the map. The zero value of the axial distance along the x axis is defined as the mid-point of the two centres of mass atoms under consideration. The maximum radial as well as axial cut-off is kept at 1.0 nm since the maximum binding of all three ions has been observed in the range of 0.5-1.0 nm in their respective PMF curves (**section III.D**). It immediately follows from FIG 2 that maximum $Na^+$ ion density is observed around 0.8 nm from the reference axis indicating weaker binding of sodium in this region as compared to $TMA^+$ and $CHO^+$ number density i.e. maximum around 0.5 nm apart from the reference axis. The highest cumulative number density has been observed for $CHO^+$ (**FIG 4.c**) which is in accordance with the study of excess number of cations at 300K (**Table 2**). We also calculate the number density of individual atoms that belong to the $TMA^+$ and $CHO^+$ molecules and it is observed that the quaternary nitrogen atom of $TMA^+$ and the oxygen atom of $CHO^+$ possess the maximum densities among all the constituent atoms in the DNA minor groove (**FIG 4.d and 4.e**).

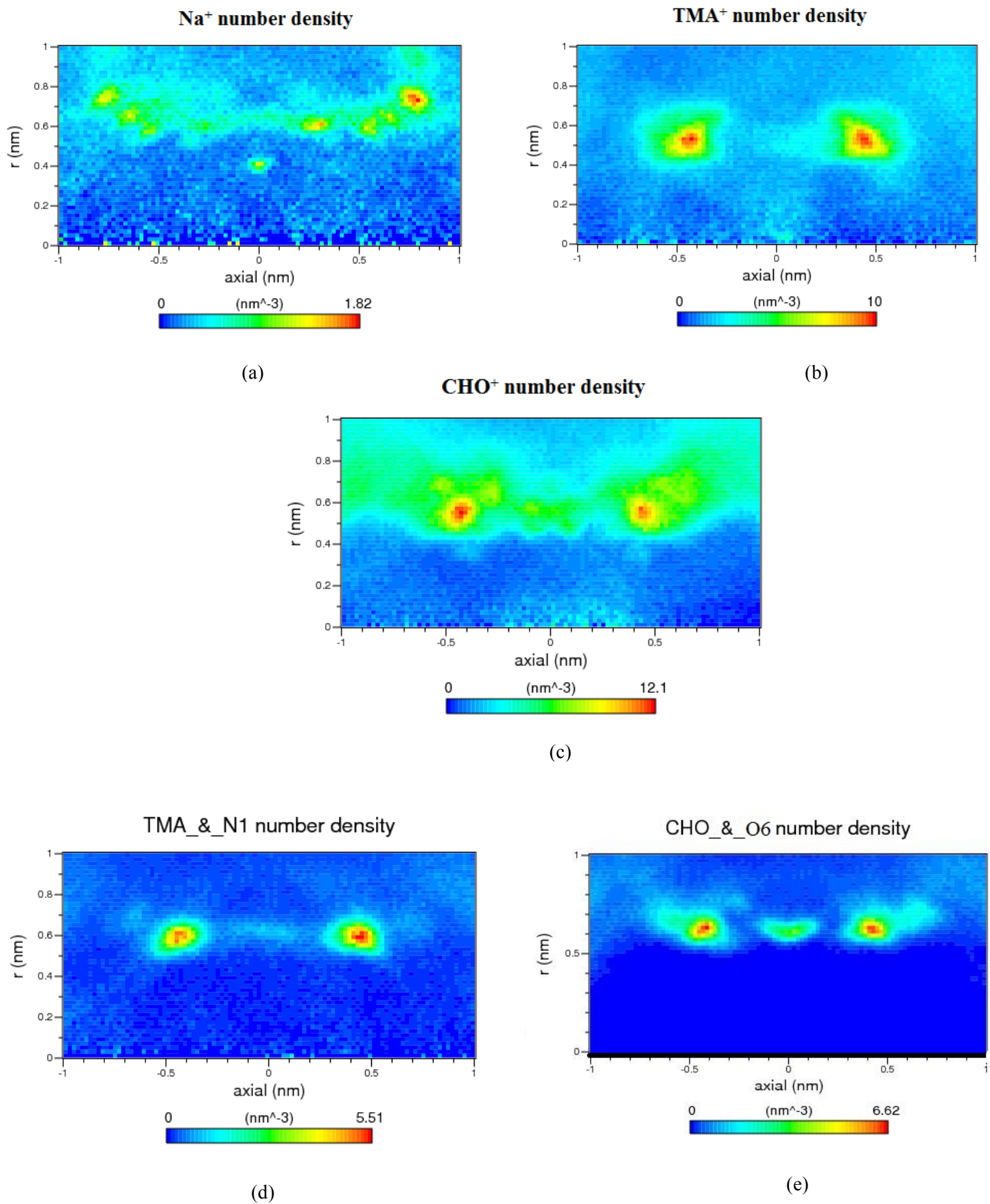

FIG 4. 2D number density maps for (a) Na$^+$ (b) TMA$^+$ and (c) CHO$^+$ bound to DNA minor groove at 300K. Nitrogen and oxygen atom number densities of (d) TMA$^+$ and (e) CHO$^+$ respectively in the DNA minor groove at 300K. The y-axis is the radial distance from the axis formed by connecting the centre of mass of atoms constituting the minor groove.

*D. Potentials of Mean Force*

Binding of small molecular and metal ions to different DNA sites through a variety of interactions may be dominated by entropy or enthalpy. This can be ascertained by studying the temperature dependence of the potentials of mean force (PMF). In the present study, PMFs are calculated using the radial distribution functions g(r) at temperature T according to eq (2).

$$W(r,T) = -k_B T \ln g(r) \qquad (2)$$

Where, g(r) is the ion-DNA segment pair correlation function, $k_B$ is the Boltzmann constant and T is the temperature of the system.

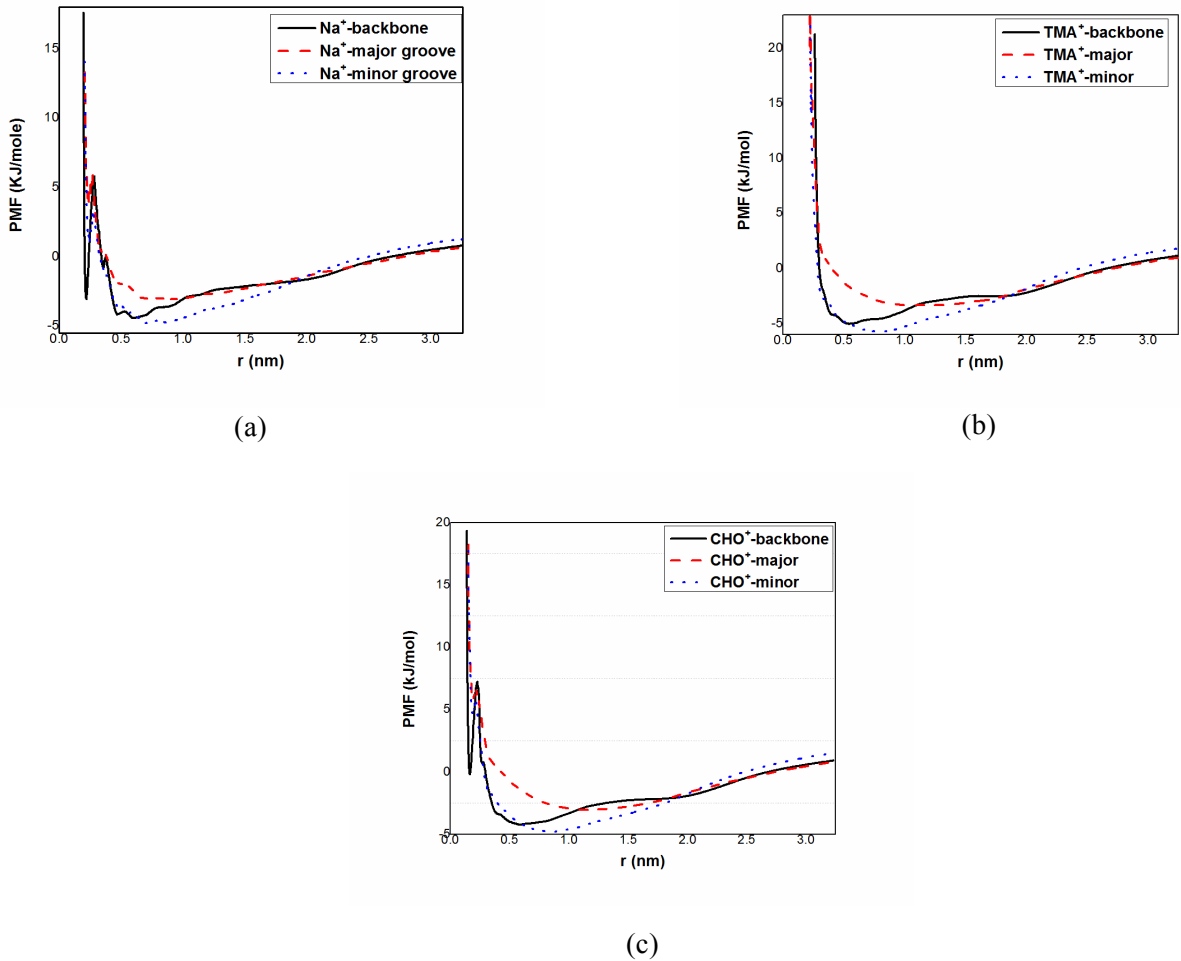

FIG 5. Potentials of mean force between (a) $Na^+$, (b) $TMA^+$, (c) $CHO^+$ and interacting DNA sites at 300K.

In order to calculate $W(r,T)$ we first calculate the radial distribution functions between the interacting DNA sites and the ionic species. PMFs are calculated between the DNA minor groove and the atom of a molecular ion having the maximum populations around the DNA minor groove (the quaternary nitrogen atom for $TMA^+$ and the oxygen atom belonging to the hydroxyl group for $CHO^+$ according to **FIG 4(d) & 4(e)**). It is to be noted that the centre of mass of the atoms constituting the DNA minor groove has not been taken into account to avoid fluctuations in the calculations due to the poor sampling induced by the flexibility of small molecular ions like $TMA^+$ and $CHO^+$. The PMFs are calculated from eq (2) between three ions and three potential DNA binding sites (backbones, major groove and minor groove) separately (**FIG 5**) at 300K temperature. The free energy of binding of any species has been found most negative for the minor groove at 300K. Furthermore, it is worth mentioning that in case of $TMA^+$ and $CHO^+$ the PMF curves exhibit the negative minima in the range of 0.5-1.0 nm (**FIG 5(b) and 5(c)**) which is in reasonable agreement with the regions of the maximum number densities of these two ions (**FIG 4(b) and 4(c)**). The integrated PMFs of ion-DNA minor groove interactions throughout the entire range of 0 to 3.2 nm for each ionic species at 300K temperature (**Table 3**) clearly indicate that the association of DNA minor groove atoms is most feasible for $CHO^+$. This is consistent with the findings on the affinity of molecular ions for DNA triplex [22]. **Table 3** contains error estimates calculated using block averaging method.

| Ions | PMF for minor groove binding (kJ/mol) at 300K |
|---|---|
| $Na^+$ | -1.650 ± 0.018 |
| $TMA^+$ | -1.721 ± 0.002 |
| $CHO^+$ | -1.810 ± 0.051 |

TABLE 3. PMF (in kJ/mol) of ion binding to DNA minor groove at 300K.

## E. Thermodynamics of Ion binding

The temperature dependent PMFs are used to calculate the enthalpy and entropy contributions to the stability of ion binding to DNA minor groove. We can decompose the entropy and enthalpy contributions as,

$$W(r,T) = \Delta H(r,T) - T\Delta S(r,T) \quad (3)$$

The entropy and enthalpy differences of the system at different distances can be derived from the following expressions [47].

$$\Delta S(r,T) = -\frac{(\partial W(r,T))}{\partial T} = -\left[\frac{W_{T+\Delta T}(r,T) - W_{T-\Delta T}(r,T)}{2\Delta T}\right] \quad (4)$$

$$\Delta H(r,T) = \left[\frac{\partial W(r,T)/T}{\partial (1/T)}\right] = W(r,T) + T\Delta S(r,T) \quad (5)$$

We calculate $W(r,T)$ at three different temperatures with $\Delta T=30K$, using which the numerical derivatives of $W(r,T)$ with respect to the temperature T are calculated. It is to be noted that the reproducible free energy values have been taken into account for consistency in the calculations.

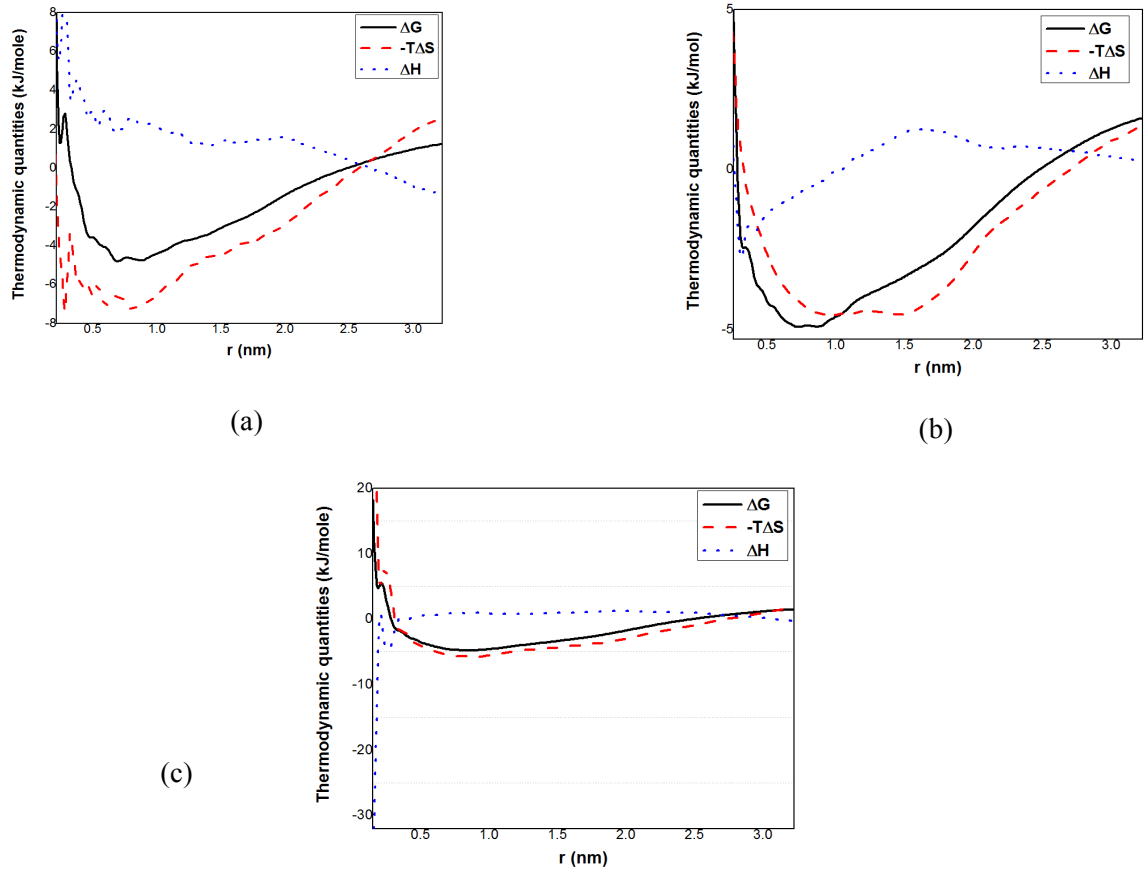

FIG 6. Thermodynamic decompositions of potentials of mean force of systems (a) $Na^+$-minor groove, (b) $TMA^+$-minor groove and (c) $CHO^+$-minor groove at 300K.

It is quite apparent from FIG 6 that in most of the cases the monovalent ion binding to the dsDNA grooves is disfavoured by enthalpy since the native hydrogen bonds between two parallel DNA helices get disrupted when hydrated molecular or metal ions penetrate the grooves [20, 21]. The ion binding processes are stabilized in all three cases mostly by entropic contributions except for $CHO^+$ where a finite enthalpic compensation is observed around 2-3 Å presumably due to additional hydrogen bonding between the hydroxyl group of $CHO^+$ and DNA minor groove atoms (**section III.G**). Interactions of sodium and tetramethyl ammonium ion with DNA minor groove atoms are solely entropy-driven at 300K (**FIG 6**).

*F. Contributions from Hydrophobic Effects*

It appears that the preferred binding sites of the molecular ions and the DNA atoms approach each other by displacing the water molecules clustering around the DNA. Therefore, we assume that hydrophobic effects are partially responsible for the large gain in entropy during the binding of ions. The entropic gain from the hydrophobic interaction can be studied by calculating the change in heat capacity for each molecular ion binding to the dsDNA minor groove since a large increase in heat capacity on solvating a hydrophobic solute in water is a defining thermodynamic signature of hydrophobic effects [48]. Understanding the thermodynamics of non-polar solvation [49] thus plays a vital role in understanding the hydrophobic interactions. We monitor the change in heat capacity of a particular ion binding to DNA minor groove as a function of the separation between the ion and the binding sites of the DNA minor groove at T= 300K using the following expression [50].

$$\Delta C_p(r,T) = -T\left(\frac{\partial^2 W(r,T)}{\partial T^2}\right)_p \quad (6)$$

$$\approx -T\left(\frac{W(r,T+\Delta T)-2W(r,T)+W(r,T-\Delta T)}{(\Delta T)^2}\right)_p \quad (7)$$

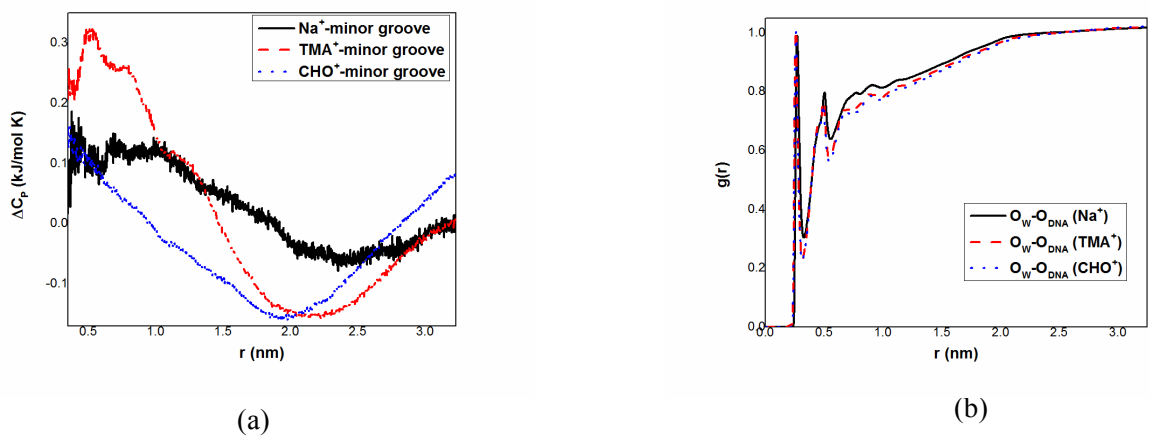

(a)  (b)

FIG 7. (a) Heat capacity as a function of the distance between ions and minor groove binding sites. (b) Pair-wise correlation function, g(r), of the water oxygen atoms as a function of distance from the oxygen atoms of the dsDNA grooves at 300K.

It immediately follows from **FIG 7** that there is a larger positive change in heat capacity for $TMA^+$-minor groove binding as compared to $Na^+$-minor groove binding. This is not unusual because the hydrophobic effects exerted by a $TMA^+$ equipped with four bulky methyl groups would obviously be higher than that of a sodium ion in solution. Furthermore, the region where the effect of hydrophobicity at 300K is most prominent is 0.5-1.0 nm (**FIG 7.a**), i.e. the region where three of the ions bind most strongly to the DNA minor groove atoms according to their PMF curves (**FIG 5**). This observation seems to be inconsistent with the molecular ion $CHO^+$ for which the variation of $C_P$ is mostly downhill in spite of its hydrophobic structural attributes. This may be rationalized by the hydrogen bond making ability of $CHO^+$ at a smaller distance from DNA minor groove atoms [51] which provides a balance between the energy gain by the formation of hydrogen bonds and the entropic profit due to the expulsion of water molecules. However among the three species, the highest magnitude of clustering of water molecules around the DNA minor groove has been observed for $Na^+$ followed by $TMA^+$ and $CHO^+$ respectively (**FIG 7.b**) which might serve as an indirect evidence for the hydrophobic signatures of $CHO^+$ not obtained from thermodynamic parameters. **FIG 7.b** signifies that the extent of preferential ion binding to the dsDNA is proportional to the depletion of the aqueous layer around it. The enthalpic contribution towards the free energy of $CHO^+$- DNA minor groove binding has been discussed in **section III.G**.

*G. Calculation of Hydrogen Bonds*

In order to investigate the persistence of the primary hydrogen bonds existing between the complementary base pairs on the two parallel DNA chains constituting the DNA minor groove in the presence of molecular or metal ions, we examine the kinetics of the different types of hydrogen bonds (H-bonds) using the g_hbond utility of GROMACS 4.5.6 [28]. It is to be noted that in addition to the native hydrogen bonds in the canonical B-DNA form, the secondary hydrogen bonds which are formed between the electronegative sites of the molecular ions and DNA minor groove atoms at 300K have also been taken into account. A geometric criterion is followed to define a hydrogen bond between a donor and an acceptor [52]. A hydrogen bond is considered being formed when the distance between the donor and acceptor is $\leq 0.35$ nm and the donor-hydrogen-acceptor bond angle is $\leq 30^0$ at 300K as obeyed in the literature [53]. In this present study we calculate the average number of hydrogen bonds formed per timeframe along with the corresponding bond length and angle distributions between all possible combinations and compute their lifetimes and relaxation times for comparison. GROMACS 4.5.6 [28] obeys the Luzar and Chandler description [54] of H-bond kinetics which considers the H-bonds labile and the lifetime of the same is calculated from the rate constants of H-bond breaking and re-forming. Since the formation of an H-bond is assumed to be an equilibrium process the free energy of H-bond formation is estimated from the corresponding equilibrium constant [55]. Lastly, the lifetime distribution of a particular kind of H-bond is converted into an autocorrelation function [56] which is fitted exponentially and then integrated to obtain the respective H-bond relaxation time [57]. The kinetic as well as the thermodynamic studies of hydrogen bonding are of particular interest to us since we assume that the disruption or the persistence of hydrogen bonds between the sites of two parallel DNA strands in the presence of a particular ion ($Na^+$, $TMA^+$ and $CHO^+$) determines the enthalpic stabilization or destabilization of the overall ion binding process to a certain extent. In **TABLE 4** we have considered only the A and T aromatic ring atoms constituting the DNA minor groove located on two parallel DNA helix which we have designated as chain A and chain B and calculated the details of hydrogen bonding between them in the presence of a particular ionic species (in parentheses). We designate the H-bonds as primary or secondary in order to distinguish between the native inter-helix hydrogen bonds and that formed between a molecular ion and DNA minor groove *in silico*. For calculating the secondary H-bonds the quaternary nitrogen atom of $TMA^+$ and the oxygen atom of $CHO^+$ have been taken into account.

| Combination | Type | No. Of H-bonds / timeframe | Lifetime (ps) | Relaxation time (ps) | Free energy of H-bond formation (kJ/mol) |
|---|---|---|---|---|---|
| Chain A-chain B ($Na^+$) | primary | 7.747±0.009 | 20.467 | 0.691 | 12.011 |
| Chain A-chain B ($TMA^+$) | primary | 7.587±0.011 | 18.001 | 0.739 | 11.693 |
| Chain A-chain B ($CHO^+$) | primary | 7.676±0.013 | 29.326 | 0.726 | 12.903 |
| Minor groove-$TMA^+$ | secondary | - | - | - | - |
| Minor groove-$CHO^+$ | secondary | 1.241±0.070 | 19.652 | 42.976 | 11.911 |

TABLE 4. Kinetics and thermodynamics of different types of hydrogen bonds at 300K.

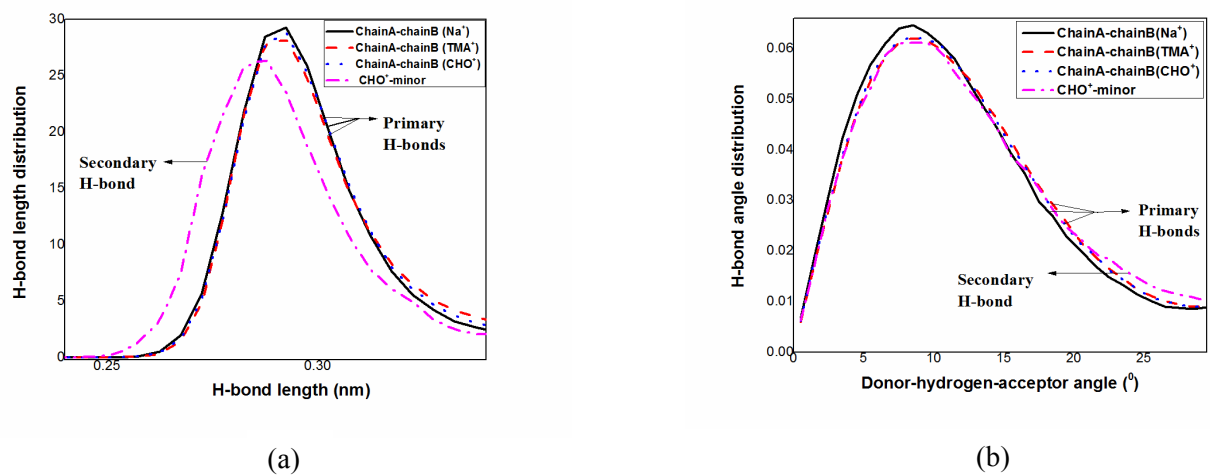

FIG 8. (a) Bond length and (b) angle distributions of primary and secondary hydrogen bonds at 300K.

It immediately follows from **TABLE 4** that the accumulation of positively charged cations around the DNA results in the disruption of the primary H-bonds between the nucleobases. The number as well as the lifetime and the energy of the primary H-bonds between chain A and chain B is reduced on replacing $Na^+$ with $TMA^+$ presumably due to the bulkier molecular ion $TMA^+$ being more effective in breaking the H-bonding network in the native DNA duplex. Parameters such as H-bond lifetime, relaxation time and free energy of formation seem to be more instrumental in determining the stability of the primary H-bonds in the presence of different ions than the numbers of H-bonds alone. The number of hydrogen bonds between an A-T base pair is 2 and thus the total number of such native H-bonds expected for the sequence $(5'-CGCGAATTCGCG-3')_2$ would be 8. In the case of $CHO^+$ however the number as well as the lifetime of the primary hydrogen bonds increases significantly along with the emergence of new secondary H-bonds between the electronegative DNA atoms and the hydrophilic –OH functional group of $CHO^+$. These secondary H-bonds, not observed for $TMA^+$ are quite stable (**TABLE 4**) and are in fact stronger than the inter-helix H-bonds of the dsDNA in the presence of positively charged ions. We observe that the secondary H-bond length distribution is shifted towards smaller H-bond length indicating the strengthening of $CHO^+$-minor groove H-bonding interactions as compared to the other primary inter-chain H-bonding in the presence of ions (**FIG 8.a**). However it has negligible impact on the angle distribution plot which indicates that secondary hydrogen bonds stabilize the $CHO^+$-minor groove interactions without affecting the angle distribution (**FIG 8.b**). The utility of the quantification of different types of hydrogen bonding is twofold. First, it supports the fact that the binding of $Na^+$ and $TMA^+$ to dsDNA are enthalpically unfavourable since the number of native H-bonds between the two helical strands are reduced significantly. Second, it justifies the substantial enthalpic contribution towards the free energy gain of DNA duplex upon binding with $CHO^+$ in the light of forming reasonably stable secondary H-bonds (**section III.E**). In addition, we also compare different types of H-bonds in terms of their free energy of formation which are indicative of the thermodynamic stabilities of both the primary as well as the secondary modes of H-bonding interactions possible in the case of $CHO^+$ (row 6, **TABLE 4**).

## IV. Summary and Conclusions

It is well-known that all the three ions reported herein are biologically relevant and are present in ambient concentrations in the human body. These ions are known to influence the overall duplex DNA stability through the negative charge neutralization of the DNA backbone and other bonding and or non-bonding interactions [57, 58] with specific sites of the DNA. According to another report [59], the concentration of NaCl is critical in determining the B to A conformational transition of the DNA where it has been shown that the DNA minor groove is occupied by sodium ions and not by the solvent molecules. The important structural roles of explicit ions in the DNA groove regions and the nature of base stacking in influencing the preference of DNA structures toward A and B conformations have been demonstrated elsewhere [60]. We hope that our work will shed lights on the thermodynamics of such preferential interactions leading to the dsDNA molecule adopting a specific conformation.

Our simulations also show that alkyl ammonium cations act differently in their association with DNA compared to monovalent alkali earth metal ions such as $Na^+$. All of the three cations under study prefer to accommodate themselves in the DNA minor groove over the major groove and DNA backbone (**FIG 3**). The calculations of PMFs determine the free energy gain in the DNA duplex upon binding with small molecular or metal ions and the PMFs further point out that the binding of $CHO^+$ to the DNA is most feasible. It is consistent with the two-dimensional number density plots of ions residing in the DNA minor groove (**FIG 4**). The calculation of temperature dependent PMFs enable us to decompose the free energy of binding for a certain ion into its entropy and enthalpy counterparts and ascertain the thermodynamic driving force for that ion binding to dsDNA. Our calculations show that apart from the small enthalpic stabilization of $CHO^+$ at smaller distances from the DNA most of the DNA-ion binding processes are favoured by entropy [61]. This is understandable since the positively charged ions are required to displace water molecules surrounding the DNA in order to access the desired binding sites on it [62]. The hydrophobic effects contribute appreciably to the entropic gain in the system upon binding of ions and dsDNA. $TMA^+$ has been found to exhibit the highest positive magnitude of $\Delta C_P$ (**FIG 7.a**).

We also focus on the change in enthalpy of binding which disfavours most of the DNA-ion binding processes. The weakening of native H-bonds between the two parallel DNA chains upon binding with ions accounts for this observation. The kinetic and distribution studies on H-bonds help reinstate the fact that positively charged ions such as $Na^+$ bind to the atoms of the phosphate groups (DNA backbone) due to strong attractive electrostatic potential, whereas $CHO^+$ ions form multiple hydrogen bonds with DNA groove atoms [23]. The free energies of the H-bonds formed between two parallel chains of the DNA in the presence of different ions correlate well with the kinetics and the bond length distributions of the same. In a nutshell, we attempt to probe the interplay of thermodynamic factors that plays a crucial role in the preferential binding of molecular and metal ions with dsDNA and provide some atomic level information about the different modes of binding possible which has not been studied in great details *hitherto* to the best of our knowledge. To understand the association of ion binding to duplex DNA even better, future studies will include the decomposition of the entropic and energetic contributions towards the PMFs into the corresponding solvent-solute, solute-solute and solvent-solvent interactions using umbrella sampling [63] or free energy perturbation [64] methods which is beyond the scope of the present work.

## ACKNOWLEDGEMENTS

We thank the super-computing facility in the department of Chemistry, Indian Institute of Technology Bombay for providing computer time for carrying out the present work**.** R.C thanks SERB **(SB/SI/PC-55/2013)**, CSIR (**Project No. 01(2781)/14/EMR-II**) and IIT Bombay-IRCC (Grant number: **12IRCCSG046)** for funding. Authors thank Prof. B. L. Tembe for stimulating discussions. S.G also thanks CSIR, Govt. of India, for a senior research fellowship.

## Author information

[*](R.C.) Telephone : + 91-022-2576 7192 Fax : + 91-022-2576 7192
 Corresponding author
[*]Email: rajarshi@chem.iitb.ac.in

**Notes**:

The authors declare no competing financial interest.

**REFRENCES**